\documentclass[twocolumn]{aastex631}

\shorttitle{Neutrino-Matter Interactions in the Cas A Supernova}
\shortauthors{Sato et al.}
\graphicspath{{./}{figures/}}

\begin{document}

\title{Examining Neutrino-Matter Interactions in the Cassiopeia A Supernova}

\correspondingauthor{Toshiki Sato}
\email{toshiki@meiji.ac.jp}

\author[0000-0001-9267-1693]{Toshiki Sato}
\affil{Department of Physics, School of Science and Technology, Meiji University, 1-1-1 Higashi Mita, Tama-ku, Kawasaki, Kanagawa 214-8571, Japan}

\author[0000-0002-8967-7063]{Takashi Yoshida}
\affil{Yukawa Institute for Theoretical Physics, Kyoto University, Kitashirakawa-Oiwakecho, Sakyo-Ku, Kyoto 606-8502, Japan}

\author[0000-0001-8338-502X]{Hideyuki Umeda}
\affil{Department of Astronomy, Graduate School of Science, University of Tokyo, 7-3-1 Hongo, Bunkyo-ku, Tokyo 113-0033, Japan}

\author[0000-0002-8816-6800]{John P. Hughes}
\affil{%
 Department of Physics and Astronomy, Rutgers University, 136 Frelinghuysen Road, Piscataway, NJ 08854-8019, USA
}%

\author[0000-0003-2611-7269]{Keiichi Maeda}
\affil{%
 Department of Astronomy, Kyoto University, Kitashirakawa-Oiwake-cho, Sakyo-ku, Kyoto 606-8502, Japan
}%

\author[0000-0002-7025-284X]{Shigehiro Nagataki}
\affiliation{%
 Astrophysical Big Bang Laboratory (ABBL), RIKEN Cluster for Pioneering Research, 2-1 Hirosawa, Wako, Saitama 351-0198, Japan
}%
\affil{%
 RIKEN Interdisciplinary Theoretical and Mathematical Science Program (iTHEMS), 2-1 Hirosawa, Wako, Saitama 351-0198, Japan
}%

\author[0000-0003-2063-381X]{Brian J. Williams}
\affil{%
 NASA, Goddard Space Flight Center, 8800 Greenbelt Road, Greenbelt, MD 20771, USA
}%

\begin{abstract}

Neutrino interactions with stellar material are widely believed to be fundamental to the explosion of massive stars. However, this important process has remained difficult to confirm observationally. We propose a new method to verify it using X-ray observations of the supernova remnant Cassiopeia A. The elemental composition in its Fe-rich ejecta that could have been produced at the innermost region of the supernova, where neutrinos are expected to interact, allows us to examine the presence of neutrino interactions. Here we demonstrate that the amount of Mn produced without neutrino nucleosynthesis processes (i.e., the $\nu$- and $\nu$p-process) is too small to explain the Mn/Fe mass ratio we measure (0.14--0.67\%). This result supports the operation of significant neutrino interactions in the Cassiopeia A supernova. If the observed Mn/Fe mass ratio purely reflects the production at the innermost region of the supernova, this would be the first robust confirmation of neutrino-matter interactions in an individual supernova. We further show that the Mn/Fe mass ratio has the potential to constrain supernova neutrino parameters (i.e., total neutrino luminosity, neutrino temperature). Future spatially-resolved, high-resolution X-ray spectroscopy will allow us to investigate the details of neutrino-supernova astrophysics through its signatures in elemental composition not only in Cassiopeia A but also in other remnants.
\end{abstract}


\section{Introduction} \label{sec:intro}

The explosion mechanism of core-collapse supernovae is a long-standing problem in astrophysics. Theoretical studies indicate that the deposition of some 1\% of the vast energy emitted by neutrinos from the collapsing stellar core can revive the stalled bounce shock and drive the supernova (SN) explosion of a massive star \citep[e.g.,][]{1985ApJ...295...14B,1995ApJ...450..830B,1996A&A...306..167J,2006A&A...447.1049B,2010PASJ...62L..49S,2012ARNPS..62..407J,2012ApJ...749...98T,2016ARNPS..66..341J,2020MNRAS.491.2715B,2021Natur.589...29B}. Therefore, observational confirmation of the neutrino interactions with stellar matter would be a major leap forward in supporting the neutrino-driven mechanism. Currently, the detection of the neutrinos from SN 1987A has been the most significant observational demonstration of this neutrino-driven mechanism \citep{1987PhRvL..58.1490H,1987PhRvL..58.1494B,1988PhLB..205..209A}. However, it is challenging to confirm and study in detail the neutrino-matter interactions only from the direct neutrino observations due to the limited number of detected events.

Neutrino interactions are expected to leave some observable signatures. One promising footprint is the elemental composition of the ejecta. For example, neutrinos can excite heavy nuclei, which subsequently emit nucleons and light nuclei, significantly altering the final composition. This process is called the ``$\nu$-process'' \citep[e.g.,][]{1988Natur.334...45W,1990ApJ...356..272W,1995ApJS..101..181W,2005PhLB..606..258H,2005PhRvL..94w1101Y,2008ApJ...672.1043Y,2013ApJ...779L...9H,2018PhRvL.121j2701H,2018ApJ...865..143S,2019ApJ...876..151S}. The $\nu$-process enhances the fraction of specific elements in the neutrino-processed ejecta; the presence of such species can therefore indicate efficient neutrino interactions. Beyond the $\nu$-process, the proton-rich environment produced by intense neutrino irradiation at the innermost region can be regarded as another piece of evidence for neutrino interactions \citep[this is known as ``$\nu$p-process'':][]{2005ApJ...623..325P,2006A&A...447.1049B,2018ApJ...852...40W}. 
In the photo-disintegrated materials, the neutrino reaction on neutrons, $n(\nu_e, e^-)p$ proceeds faster than on protons $p(\bar{\nu}_e, e^+)n$ (due to neutron's larger mass) \citep[e.g.,][]{2005ApJ...623..325P,2006ApJ...637..415F,2006PhRvL..96n2502F}.
Thus, in an environment where the $\nu_e$ and $\bar{\nu_e}$ spectra are similar, as expected in the early neutrino wind from the proto-neutron star, the weak-force processes drive the ejecta to a proton-rich composition. 
As with the $\nu$-process, such proton-rich ejecta can be identified by their elemental composition. Therefore, investigating the elemental composition of the neutrino-processed ejecta will provide us a unique opportunity to test the extent of neutrino interactions in the exploding massive star. On the other hand, until recently, it was difficult to observe the innermost region of an exploding star separately from the outer layers.

\begin{figure*}[t!]
 \begin{center}
  \includegraphics[bb=0 0 1472 482, width=16cm]{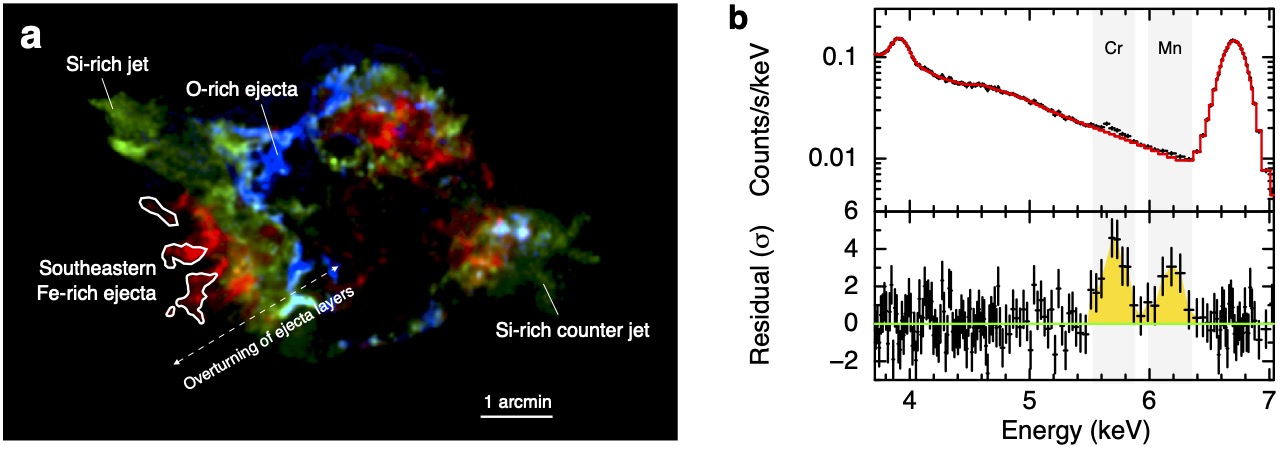}
 \end{center}
\caption{(a) Asymmetric ejecta distribution in the Cassiopeia A supernova remnant. The red and blue colors highlight Fe-rich and O-rich emission regions, respectively. The {\it Chandra} ratio map of the Si/Mg band is shown in green color where the jet-like structures can be seen at the northeastern and southwestern directions \citep[e.g.,][]{2004ApJ...615L.117H,2006ApJ...644..260L,2022PASJ...74..334I}. In the southeastern region, it has been proposed \citep{2000ApJ...528L.109H,2017ApJ...842...13W,2021A&A...645A..66O,2020casATi} that the apparent overturning of ejecta layers was produced by plumes from rising bubbles in the neutrino-driven convection layer during the supernova explosion. (b) X-ray spectrum (black data) from the Fe-rich region. The red curve shows the best-fit thermal model without Cr and Mn emissions. In the 5.5--5.9 keV and 6.0--6.3 keV photon energy bands (denoted with faint gray and yellow regions), X-ray line features from shocked Cr and Mn are clearly identified. }
\label{fig:CasA}
\end{figure*}

Observations of elements in asymmetric supernova remnants can provide unique information on the innermost regions of core-collapse supernovae \citep[e.g.,][]{1994A&A...284L...1I,2001ApJ...560L..79V,2006ApJ...647L..41R,2011ApJ...732..114L,2012Natur.490..373G,2014Natur.506..339G,2017ApJ...834...19G,2015Sci...347..526M,2017ApJ...844...84H,2018ApJ...856...18K,2018SSRv..214...44L,2019ApJ...878L..19T,2020ApJ...904..115L,2020ApJ...889..144H,2021A&A...646A..82P}. Especially, the Fe-rich ejecta located at the southeastern region of Cassiopeia A observed in X-rays (Fig.~\ref{fig:CasA}a) are thought to have been produced at the innermost region of the supernova explosion \citep[][]{2000ApJ...528L.109H,2000ApJ...537L.119H,2002A&A...381.1039W,2004NewAR..48...61V,2003ApJ...597..362H,2004ApJ...615L.117H,2009ApJ...703..883H,2012ApJ...746..130H,2017ApJ...837..118L,2020casATi}. Important pioneering studies of the Fe-rich ejecta have been done by \cite{2003ApJ...597..362H,2009ApJ...703..883H,2012ApJ...746..130H}, where the plasma properties and the observed elemental compositions in knotty Fe-rich structures support the idea that this region experienced complete Si burning with $\alpha$-rich freeze out during the explosion. It has been proposed that the overturning of ejecta layers in this part of the remnant was driven by hot bouyant bubbles produced in the neutrino-driven convection layer during the supernova explosion \citep{2000ApJ...528L.109H,2016ApJ...822...22O,2017ApJ...842...13W,2021A&A...645A..66O,2020casATi,2022ApJ...932...93T}. This structure, therefore, is ideal for studying the inner workings of the engine that drives the explosions of massive stars. \cite{2020casATi} have measured the mass ratios among Ti, Cr and Fe in the Fe-rich ejecta, which supports a major role of a high-entropy process for driving the explosion of the Cassiopeia A supernova. Currently, such abundance measurements provide the only way to constrain local physical parameters (i.e., lepton fraction, peak temperature, peak density and so on) associated with the neutrino-driven SN engine. The comparison of the elemental mass ratios with those from theoretical calculations supports that the structure was produced at the strongly neutrino-processed proton-rich region. The high-entropy Fe-rich plumes appear to connect smoothly with the characteristic three-dimensional structures (i.e., bubble-like interior and outer ring-like structures) of the ejecta in Cassiopeia A \citep[e.g.,][]{2010ApJ...725.2038D,2013ApJ...772..134M,2015Sci...347..526M,2014MNRAS.441.2996A}, and these observed features have been reproduced well in recent multi-dimensional simulations for this remnant \citep{2016ApJ...822...22O,2017ApJ...842...13W,2021A&A...645A..66O}.

In this paper, we propose a new method to verify the neutrino interactions in the Cassiopeia A supernova using the elemental composition in the Fe-rich ejecta. Here we focus on the production of Mn which is enhanced by neutrino effects \citep[e.g.,][]{1990ApJ...356..272W,2008ApJ...672.1043Y}. In the $\nu$-process, neutrino spallation, working on the $^{56}$Ni produced by complete and incomplete Si burning, enhances the Mn abundance through the reaction\footnote{Neutral-current reactions like this one, which do not depend on neutrino flavor, are the main contributor to the $\nu$-process, although there is also some contribution from charged-current reactions.} $^{56}{\rm Ni}(\nu,\nu^\prime p)^{55}{\rm Co}$, where $^{55}$Co decays to $^{55}$Fe and then to stable $^{55}$Mn. The timescale for the production of $^{56}$Ni after the SN shock passage is on the order of 0.1 s. In the subsequent few seconds, $^{55}$Co is produced via neutrino irradiation. Especially in the complete Si burning regime, the production of Mn without the $\nu$-process is much smaller than with it \citep[as we show below, and see also][]{2008ApJ...686..448Y}. In addition to the $\nu$-process, the strongly neutrino-processed proton-rich environment can also enhance the Mn abundance at the complete Si burning region \citep{2018ApJ...852...40W}, where $^{55}$Mn is initially produced as the proton-rich isotope $^{55}$Ni, which then decays via Co and Fe to $^{55}$Mn. In this case, the innermost material is strongly irradiated by neutrinos during the accretion phase, producing a proton-rich environment that synthesizes $^{55}$Ni effectively. After the accretion phase and the SN shock passage, neutrinos still irradiate the central materials during the cooling phase, so the $^{56}{\rm Ni}(\nu,\nu^\prime p)^{55}{\rm Co}$ reaction increases the amount of Mn further. Thus, both the $\nu$-process and a proton-rich environment (i.e., the $\nu$p-process) can enhance Mn production, which means that we can test the existence of the neutrino interaction by measuring the Mn abundance in the innermost ejecta. 

\section{Observation of the Fe-rich ejecta in Cassiopeia A} \label{sec:obs}
Cassiopeia A is one of the most well-studied young Galactic SNRs with an age of $\sim$350 years \citep{2001AJ....122..297T,2006ApJ...645..283F}. The remnant has been observed with Chandra X-ray imager ACIS-S (Advanced CCD Imaging Spectrometer) several times since the launch \citep[e.g.,][]{2000ApJ...528L.109H,2000ApJ...537L.119H,2004ApJ...615L.117H,2011ApJ...729L..28P,2012ApJ...746..130H,2014ApJ...789..138P,2017ApJ...836..225S,2018ApJ...853...46S}. We used all the ACIS-S data targeting the Cassiopeia A from 2000 to 2018, with a total exposure of around 1.57 Ms. We reprocessed the event files (from level 1 to level 2) to remove pixel randomization and to correct for charge-coupled device (CCD) charge transfer efficiencies using CIAO version 4.6 and CalDB 4.6.3. The bad grades were filtered out, and good time intervals were reserved.

Figure \ref{fig:CasA}a shows a three-color image of Cassiopeia A obtained by Chandra in 2004, where the Fe-rich, O-rich and Si-rich emissions are highlighted in red, blue and green, respectively. We extracted the X-ray spectrum in 3.7--7.1 keV (Figure \ref{fig:CasA}b) from the Fe-rich region defined in Figure \ref{fig:CasA}a and fitted it with an absorbed thermal plasma model with a gsmooth model that is used for modelling the lines broadened by thermal broadening and/or multiple velocity components (= phabs $\times$ gsmooth $\times$ vvpshock in Xspec using AtomDB version 3.0.9). The spectral analysis revealed a clear Mn-K$\alpha$ line feature at $\sim$6.0--6.3 keV (Figure \ref{fig:CasA}b) at a confidence level of 5.1$\sigma$ with a measured Mn/Fe mass ratio of 0.14--0.67\% ($\pm$99\% confidence range). For more details on this analysis, we refer to our previous study \citep[see Methods in][and Fig.~\ref{fig:CasA}b]{2020casATi}. To check the robustness of the measurements, we also estimated the Mn abundances using different observational epochs (2004 versus all the other data) and found that the Mn/Fe mass ratios are 0.10--0.70\% (99\% confidence level) in 2004 and 0.08--0.97\% (99\% confidence level) in the other 13 epochs from 2000 to 2018, showing good agreement. Even if we used the independent SPEX analysis code (version 3.06.01), the estimated Mn/Fe mass ratio becomes 0.30--1.07\% (99\% confidence level) using all the Chandra data, which is consistent with that from Xspec (AtomDB 3.0.9). 

We note that the current spectral modeling in a wide energy band up to $\sim$10 keV has an issue around 7.1--8.5 keV. To reduce this effect, we used the result from the narrow band fitting in 3.7--7.1 keV in this research. This issue seems to come from the uncertainty on the Fe K$\beta, \gamma$, and higher energy emissions, which makes modeling the continuum emission difficult. The uncertainty on the modeling of the continuum emission could affect the measurement of weak emission lines such as Mn. On the other hand, we found the observed Mn/Fe ratios are consistent with each other even when the fitting range was extended \citep[see Methods in][]{2020casATi}. Still, continuum modeling is a potential source of systematic uncertainty on measuring the abundances of trace Fe-group elements like Ti, Mn, and Cr. Future X-ray observations with X-ray calorimeter missions, such as XRISM \citep{2018SPIE10699E..22T}, will provide more robust composition measurements.

\subsection{Contamination effect}

In this study, we assume that the observed Mn purely reflects production at the innermost region of the SN explosion because the low Mn/Fe and Cr/Fe ratios can rule out its origin in the quasi-statistical equilibrium region \citep{2020casATi}. Not only that, in the Fe-rich region, \cite{2003ApJ...597..362H,2012ApJ...746..130H} have also found a knot where the 0.8--8 keV X-ray spectrum comes from almost pure Fe. Furthermore, recent near-infrared observations have provided additional evidence for the presence of dense knots formed through complete silicon burning with $\alpha$-rich freezeout within the plume region \citep{2023arXiv230504484K}. Therefore, there is no doubt about the existence of such a plasma in the region, which consists of the innermost supernova material. Here, we note that contamination from the overlaying Si-rich region (i.e., the incomplete Si burning region) could contribute to the emissions of Ti, Cr and Mn in this region. However, \cite{2020casATi} found that the abundance pattern of the Si-rich component potentially contaminating the Fe-rich region does not explain the amount of these elements, which suggests the contamination is negligibly small. In addition, the absence of pileup peaks from strong Si and S lines in the 3.7--7.1 keV range spectrum \citep[see][about the pileup peaks]{2022ApJ...932...93T} suggests that the contribution of Si-rich ejecta in this region is small. The Fe-rich region contains a strong line of Ni at 7.8 keV, and the  Ni/Fe abundance ratio supports the identification of $\alpha$-rich freeze-out material \citep[see][and Figure \ref{fig:ap_f2}a]{2020casATi}.

If we assume an extreme case that all the emissions of Ti, Cr and Mn are due to contamination from the Si-rich ejecta (i.e., the production at the incomplete Si burning layer), it would cause some nucleosynthetic difficulties. The biggest problem is that the synthesized amount of Ti and Cr in the $\alpha$-rich freeze-out would have to be unrealistically small in this case. Based on the Chandra observations, \cite{2022ApJ...932...93T}, find that at least 80\% of the Fe in this region\footnote{Tsuchioka et al.'s 80\% estimate is for the southeast Fe-rich region. \cite{2012ApJ...746..130H} obtained a global estimate that about 25\% of the Fe in Cas A is produced in $\alpha$-rich freeze out. Hwang \& Laming's value is expected to be slightly higher in the SE Fe-rich region. To be conservative we consider the 80\% value to be an upper bound on the $\alpha$-rich freeze-out fraction in the SE region.} must have been produced by $\alpha$-rich freeze-out. A complete absence of Ti and Cr in the $\alpha$-rich freezeout region would therefore be unacceptable from the nucleosynthetic perspective of core-collapse supernovae. In summary, for the Si-rich contamination scenario to be a viable explanation for the Fe-group mass ratios, one or more of the following unlikely issues would need to be significant: the observations are invalid due to some unknown instrumental effects; the relative mass ratios derived from the observations are invalid due to some error in the atomic data; or the predicted nucleosynthetic yields in the Si burning layers need to be completely revised. To check the first point (instrumental effects), we also analyzed the data obtained by other satellites (XMM-Newton and Suzaku), which supports our results (see Appendix \ref{sec:other}). However, deeper and more precise observations will be needed for more reliable verification.

\begin{figure*}[t!]
 \begin{center}
  \includegraphics[bb=0 0 1252 902, width=16cm]{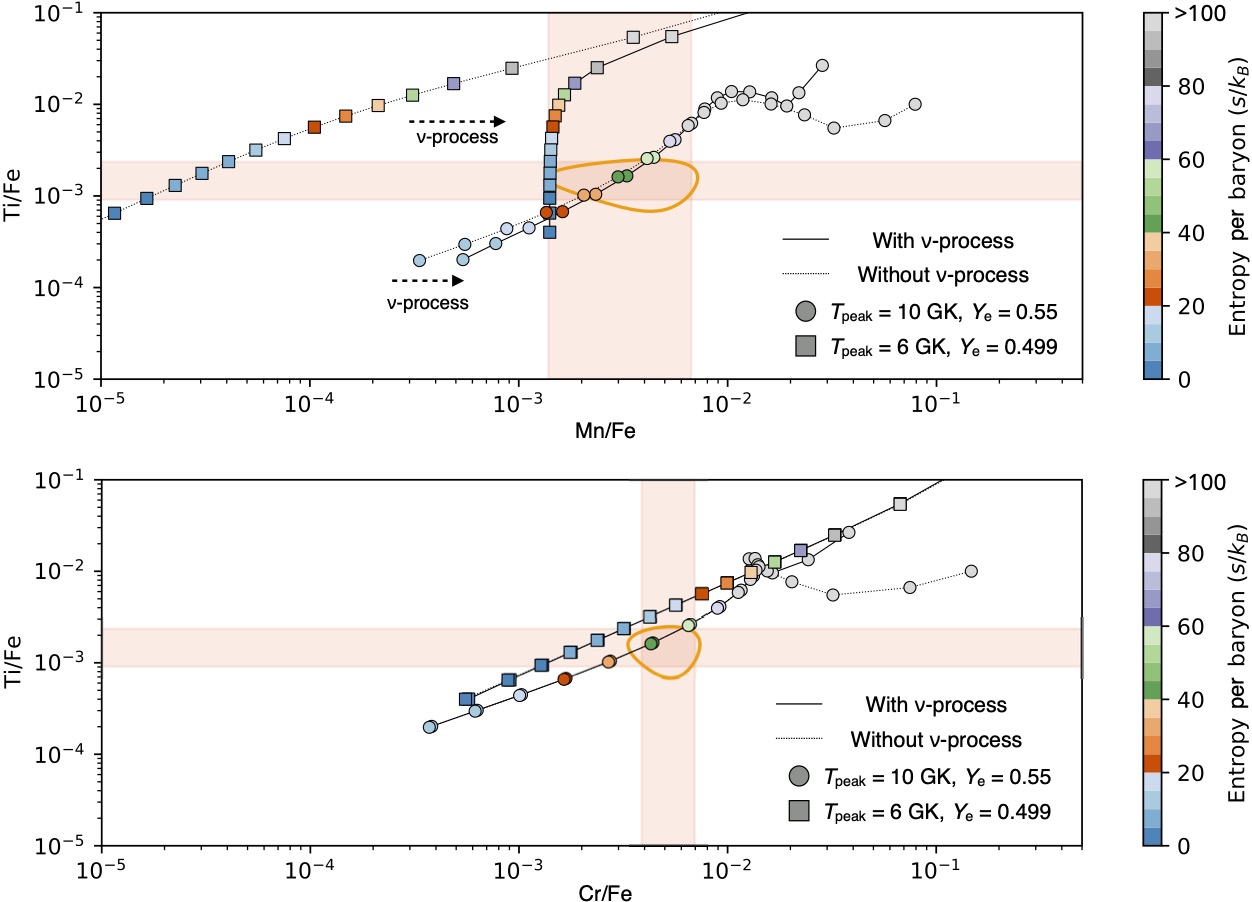}
 \end{center}
\caption{The Ti/Fe, Mn/Fe (top) and Cr/Fe (bottom) mass ratios observed in the Fe-rich ejecta. The faint orange areas are the combination of two of the mass ratios (99\% confidence level, $\Delta\chi^2$ = 6.64). The orange lines indicate the 99\% confidence contour for two parameters ($\Delta\chi^2$ = 9.21). The colored points show the mass ratios in our nucleosynthesis calculations. The color scale indicates the entropy per baryon at the peak temperature in each calculation. The solid and dotted lines indicate the calculations with and without the $\nu$-process, respectively. Here the total neutrino energy is assumed to be 3$\times$10$^{53}$ erg, which corresponds to the binding energy for a neutron star with $\sim$1.4 $M_\odot$. The curves plotted with square data points show a parameter study for the typical $\alpha$-rich freeze-out environment ($T_{\rm peak} =$ 6 GK, $Y_{\rm e}$ = 0.499), while changing the initial density from 10$^{5.5}$ g cm$^{-3}$ to 10$^{7.5}$ g cm$^{-3}$ (Cases a and a$^\prime$). The curves plotted with circle data points show the same, but for hot ($T_{\rm peak} =$ 10 GK) and proton-rich ($Y_{\rm e}$ = 0.55) environment (Cases b and b$^\prime$). In the bottom panel the curves with and without the $\nu$-process lie mostly on top of each other, except for the proton-rich environment case (circle symbols) at the largest entropy values (which are, anyway, far from the allowed observational range).}
\label{fig:f1}
\end{figure*}

\section{Neutrino interactions verified by measuring Manganese} \label{sec:mn}

To verify the role of neutrino interactions during the SN explosion, we compare the observed Mn/Fe mass ratio of 0.14--0.67\% (99\% confidence level) with theoretical calculations including neutrino effects (Fig.~\ref{fig:f1}). Here we consider several simple cases to test the neutrino interaction hypothesis:
\begin{description}
   \item[Case a]\mbox{} Assumes slightly neutron-rich ejecta for the typical $\alpha$-rich freeze-out environment ($Y_{\rm e}$ = 0.499, $T_{\rm peak} =$ 6 GK) with the $\nu$-process, where the neutron-rich environment is realized by weak interactions (e.g., electron capture) during the progenitor evolution.
   \item[Case a$^{\prime}$]\mbox{} The same as case a, but not including the $\nu$-process.
   \item[Case b]\mbox{} Assumes proton-rich ejecta ($Y_{\rm e}$ = 0.55, $T_{\rm peak} =$ 10 GK) with the $\nu$-process, where the proton-rich environment is produced by the
   $\nu p$-process operating during the earlier accretion phase of the explosion.
   \item[Case b$^{\prime}$]\mbox{} The same as case b, again not including the $\nu$-process.
\end{description}
The elemental compositions at each SN environment are calculated separately for a range of initial density values from 10$^{5.5}$ g cm$^{-3}$ to 10$^{7.5}$ g cm$^{-3}$. For the calculations we utilize the thermodynamic trajectory taken from a 1D SN model \citep[see Methods in][for more details]{2020casATi}. We set neutrino temperatures of $T_{\nu_{\rm e}} = T_{\bar{\nu}_{\rm e}} =$ 4 MeV/$k$ and $T_{\nu_{\rm x}} =$ 6 MeV/$k$ for the $\nu$-process, where we assume Fermi-Dirac distributions with zero chemical potential for the neutrino energy spectra. Here, the neutrino flux decays exponentially with a timescale of 3 s. We used the $\nu$-process cross sections for $^{56}$Ni studied in \cite{2009PhRvC..79f1603S}, $^4$He and $^{12}$C studied in \cite{2008ApJ...686..448Y}. For the others, the cross sections evaluated in \cite{1990ApJ...356..272W} were used (the data are summarized in Hoffman \& Woosley 1992\footnote{\url{http://dbserv.pnpi.spb.ru/elbib/tablisot/toi98/www/astro/hw92\_1.htm}}). The nucleosynthesis calculations used in this study have been constructed in our previous studies \citep{2020ApJ...893...49S,2020casATi}. Through this analysis, we conclude that the observed Mn/Fe mass ratio can not be reproduced without neutrino interactions (Fig.~\ref{fig:f1} top). We need either the $\nu$-process or a neutrino-heated proton-rich environment (Case a, b or b$^{\prime}$) to reproduce the observed mass ratios, with the proton-rich cases in better agreement with the observations. The amount of Mn produced at the complete Si burning layer with slightly neutron-rich ejecta including the $\nu$-process (case a, Fig. \ref{fig:f1} top panel box symbols) is just marginally consistent with the observed mass ratio. The case without the $\nu p$- and $\nu$-process (case a$^{\prime}$) yields almost two orders of magnitude less Mn, far from the observed value. 

The production of $\alpha$ elements like Ti and Cr is not so sensitive to neutrino interactions but is sensitive to the entropy and the rate of change in temperature and density \citep[Fig.~\ref{fig:f1} bottom, also see][]{1973ApJS...26..231W,2020casATi}. The mass ratios calculated from the neutron-rich ejecta with $Y_{\rm e} = 0.499$ (the square data points in Fig.~\ref{fig:f1} bottom) are slightly out of the observational range due to efficient $^{48}$Cr production (decays to stable $^{48}$Ti) on the neutron-rich side \citep[see Fig.~10 in][]{2018ApJ...852...40W}. Therefore, current measurements prefer neutrino-heated proton-rich ejecta as the origin of the Fe-rich ejecta in agreement with our recent previous study \citep{2020casATi}. Here the observed mass ratios indicate that the entropy per baryon is in the range of $\sim$32--58 $k_B$, which is in rough agreement with that of the high-entropy proton-rich ejecta in multi-dimensional simulations \citep[e.g.,][]{2018ApJ...852...40W,2018MNRAS.477.3091V}. We note that different thermodynamic trajectories could change this estimation.

The circle data points in Fig.~\ref{fig:f1} show the results for the strongly neutrino-heated proton-rich ejecta ($T_{\rm peak} = 10$ GK). In this case, the nucleosynthesis calculations either with or without the $\nu$-process agree with the observed Mn/Fe ratio. Here, Mn is efficiently produced via $^{55}$Ni \citep{2018ApJ...852...40W}; thus Mn production via the $\nu$-process is sub-dominant.

\begin{figure*}[t!]
 \begin{center}
  \includegraphics[bb=0 0 1276 546, width=16cm]{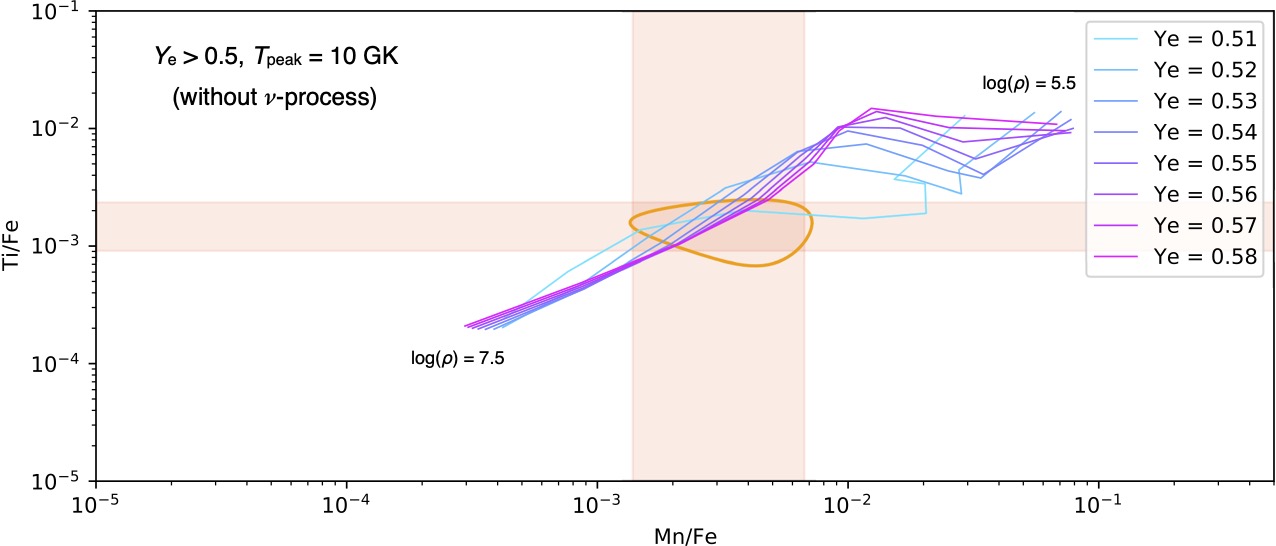}
 \end{center}
\caption{The Ti/Fe and Mn/Fe mass ratios observed in the Fe-rich ejecta (same as the top panel of Fig.~\ref{fig:f1}) comparing with models that have values of $Y_{\rm e}$ varying from 0.51 to 0.58. The coloured curves show a parameter study for hot and proton-rich ejecta ($T_{\rm peak} =$ 10 GK, $Y_{\rm e} >$  0.5), while changing the initial density from 10$^{5.5}$ g cm$^{-3}$ to 10$^{7.5}$ g cm$^{-3}$. We used the thermodynamic trajectory taken from the innermost region of our 1D SN model ($Z = 0.5 Z_\odot$ and $E_{\rm exp} =$ 3$\times$10$^{51}$ erg) to calculate these models. The high explosion energy for the Cassiopeia A supernova has been supported by some previous studies \citep[e.g.,][]{2012ApJ...746..130H,2020ApJ...904..115L,2020ApJ...893...49S}
}.
\label{fig:diffYe}
\end{figure*}

The proton-rich environment must have been realized before $^{56}$Ni is formed (i.e., during the accretion phase). If the hot material produced in such an environment comes out early as hot plumes or neutrino-driven winds, the timescale of neutrino irradiation may be reduced. This effect could reduce the effect of the $\nu$-process on the proton-rich ejecta. On the other hand, as seen
in Fig.~\ref{fig:f1}, the difference between the cases with and without  the $\nu$-process is very small. Thus, in the absence of a specific model for early lift-off, we tentatively conclude that this effect 
is likely to be negligible. 

In the proton-rich ejecta produced in multidimentional simulations, the electron fraction shows a wide distribution of $Y_{\rm e} =$ 0.51--0.58 \citep[e.g.,][]{2017ApJ...843....2H,2018ApJ...852...40W,2018MNRAS.477.3091V,2020MNRAS.491.2715B}. The electron fraction in the nuclear burning region changes the production efficiency of $^{55}$Ni (= $^{55}$Mn). In Figure \ref{fig:diffYe}, we investigate this effect. We find that the Mn/Fe mass ratio gradually changes with the electron fraction within the range of the observed value, which means that the abundance ratio has the potential to limit the electron fraction in the proton-rich ejecta. If a more precise measurement of Mn (and also Ni) becomes feasible, it will also be possible to estimate $Y_{\rm e}$ independently of entropy estimates from Cr and Ti. For example, multi-dimensional simulations suggest a positive correlation between the entropy and electron fraction in the neutrino-processed ejecta \citep[e.g.,][]{2018ApJ...852...40W,2018MNRAS.477.3091V}, which will be able to be tested by future X-ray observations of supernova remnants.

\begin{figure*}[t!]
 \begin{center}
  \includegraphics[bb=0 0 1162 556, width=16cm]{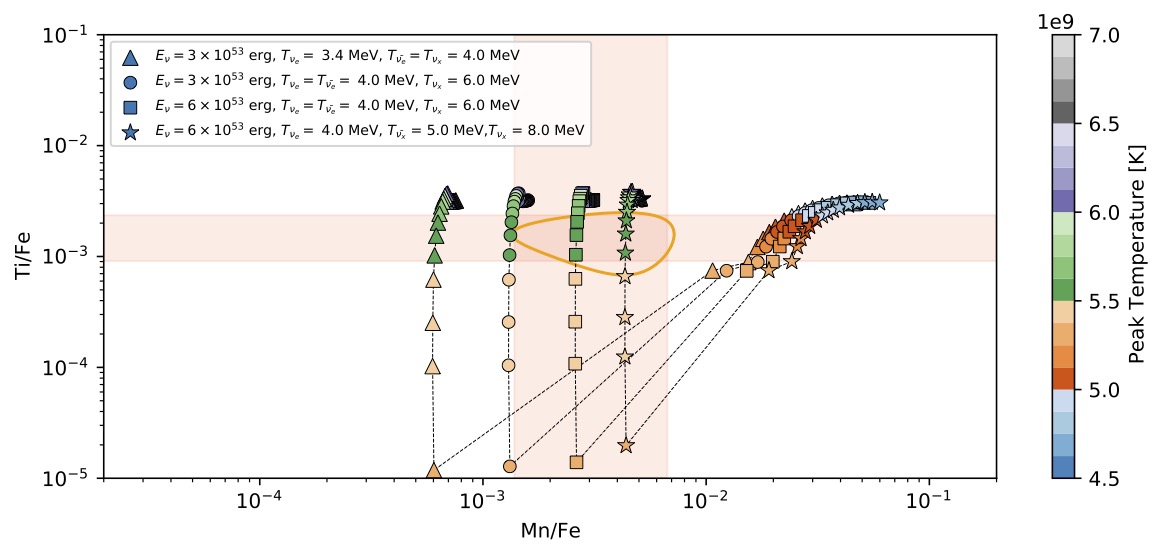}
 \end{center}
\caption{The Ti/Fe and Mn/Fe mass ratios (same as in Fig.~\ref{fig:f1}) computed for supernova models that have different neutrino parameters. The colored points show the mass ratios in our nucleosynthesis calculations for core-collapse SNe. The color scale indicates the peak temperature in the models. We assumed a 15 $M_{\odot}$ progenitor with the sub-solar metallicity \cite{2020ApJ...893...49S} of $Z = 0.5 Z_{\odot}$ and a high explosion energy of 3$\times$10$^{51}$ erg \citep[see also][for arguments supporting the high explosion energy for the Cassiopeia A supernova]{2012ApJ...746..130H,2020ApJ...904..115L}. Different total neutrino energies (3$\times$10$^{53}$ erg and 6$\times$10$^{53}$ erg) and neutrino temperatures are assumed for the curves plotted with triangle, circle, square and star data points. The observed ratios (colored area) are shown for demonstrating how useful these can be in limiting neutrino parameters, with the caveat that the analysis here is not intended to explain these ratios. }
\label{fig:f2}
\end{figure*}

\section{Different neutrino parameters for $\nu$-process} \label{sec:mn2}

As discussed above, a proton-rich environment in the neutrino-heated bubble best describes the measurements of the Cr, Ti, Mn, and Fe abundances from current X-ray observations. On the other hand, distinguishing between the two scenarios (proton-rich nucleosynthesis and $\nu$-process) using the current observational data is not straightforward \citep[see also][]{2020casATi}. Also, we note that it is possible the Fe-rich ejecta are a mixture of proton-rich ejecta and $\nu$-process ejecta. For example, a more finely-detailed analysis performed for the southeastern Fe-rich region suggests that the abundance ratios could vary from region to region (see Appendix \ref{sec:detail}), implying different physical conditions. To test these, we would need spatially-resolved high-resolution X-ray spectroscopy realized by advanced X-ray observatories, such as Athena \citep{2018SPIE10699E..1GB} and Lynx \citep{2019JATIS...5b1001G}, which are planned for the 2030's and beyond. Such analysis will potentially lead to different abundance patterns at different regions, with different contributions of $\nu$- and $\nu p$- processes. Therefore, despite our present conclusion that the `entire' Fe-rich region predominantly experienced the $\nu p$-process, we investigate in the subsequent analysis how the contribution of the $\nu$-process (alone) can affect the composition of the Fe-rich ejecta. The goal here is to give insight into these questions with an eye toward future X-ray missions.

In case of the $\nu$-process, the synthesized amount of Mn depends on the total neutrino luminosity and the neutrino temperature during the explosion \citep[e.g.,][]{2005PhRvL..94w1101Y,2008ApJ...686..448Y}. Here we examine how these affect our interpretation, using 1D SN models that include more realistic trajectories (Fig.~\ref{fig:f2}) than adopted previously (Fig.~\ref{fig:f1}). 

The total energy carried by neutrinos is mainly determined by the binding energy of the neutron star that is formed by the supernova \citep{2001ApJ...550..426L}. Thus, we can roughly estimate the total neutrino energy $E_{\nu}$ as
\begin{equation}
    E_{\nu} \approx E_{\rm binding} \approx 1.5 \times 10^{53} ~(M/M_\odot)^2~{\rm erg}.
\end{equation}
The mass of Cassiopeia A's neutron star has been estimated to be $\approx$1.3--2.0 $M_\odot$ \citep{2011MNRAS.411.1977Y}, so the neutrino energy corresponds to (2.5--6)$\times 10^{53}$ erg. In Fig.~\ref{fig:f2}, we show the mass ratios computed for the 1D SN models with two different neutrino energies (3$\times 10^{53}$ erg and 6$\times 10^{53}$ erg) while also changing neutrino temperatures. We set the neutrino temperatures based on previous studies \citep[e.g.,][]{1990ApJ...356..272W,1995ApJS..101..181W,2005PhRvL..94w1101Y,2005PhLB..606..258H,2019ApJ...876..151S}. 

We see a clear tendency that the higher the luminosity and temperature, the greater the production of Mn. The former provides a potential measure of the nature of the compact object as discussed above, and the latter may be useful to constrain the properties of neutrino emissions as related to some uncertainties in the theoretical modeling. For example, a recent study investigated the $\nu$-process using time-dependent neutrino emission spectra \citep{2019ApJ...876..151S}, where the average energy of the neutrinos seems to be noticeably lower than previous studies \citep[e.g.,][]{1990ApJ...356..272W,1995ApJS..101..181W,2005PhRvL..94w1101Y,2005PhLB..606..258H}. In addition, neutrino flavor oscillations, whose effects are omitted here, could produce more complicated neutrino spectral shapes and could also change the $\nu$-process yields \citep{2006ApJ...649..319Y,2008ApJ...686..448Y,2015PhRvD..91f5016W,2020ApJ...900..144X}. While further investigations into these effects lie outside the scope of this paper, the analysis here demonstrates the potential power of the detailed nucleosynthesis study to provide strong constraints on the neutrino spectrum. 

We note that the amount of $\nu$-process elements depends strongly on where the materials were exposed to neutrinos. In other words, moving the mass-cut position changes the amount of Mn in our 1D calculations. In our cases, we determined the location of the mass cut so that the ejected $^{56}$Ni mass is 0.07 $M_\odot$ \citep[see][]{2020ApJ...893...49S}, which mimics a typical core-collapse SN (such as SN 1987A) as the simplest assumption\footnote{\cite{2012ApJ...746..130H} have estimated the Fe mass of $\sim$0.09--0.13 $M_\odot$ in X-rays, depending on assumptions about the ejecta filling factor. While this value is slightly larger than our assumptions, the total amount of Fe is not physically significant in our calculations as the mass-cut position is given by hand.}. Therefore, if we assume a smaller explosion energy, the mass-cut position moves inward and the amount of Mn via the $\nu$-process increases (see Appendix \ref{sec:model}). In multi-dimensional cases, this situation would be more complicated. Ideally, an argument using multiple elements related to neutrino processes, not just Mn in this case, would lead to more accurate limits on neutrino parameters.

\section{Summary and Future prospects} \label{sec:summary}

\begin{figure*}[t!]
 \begin{center}
  \includegraphics[bb=0 0 1492 770, width=16cm]{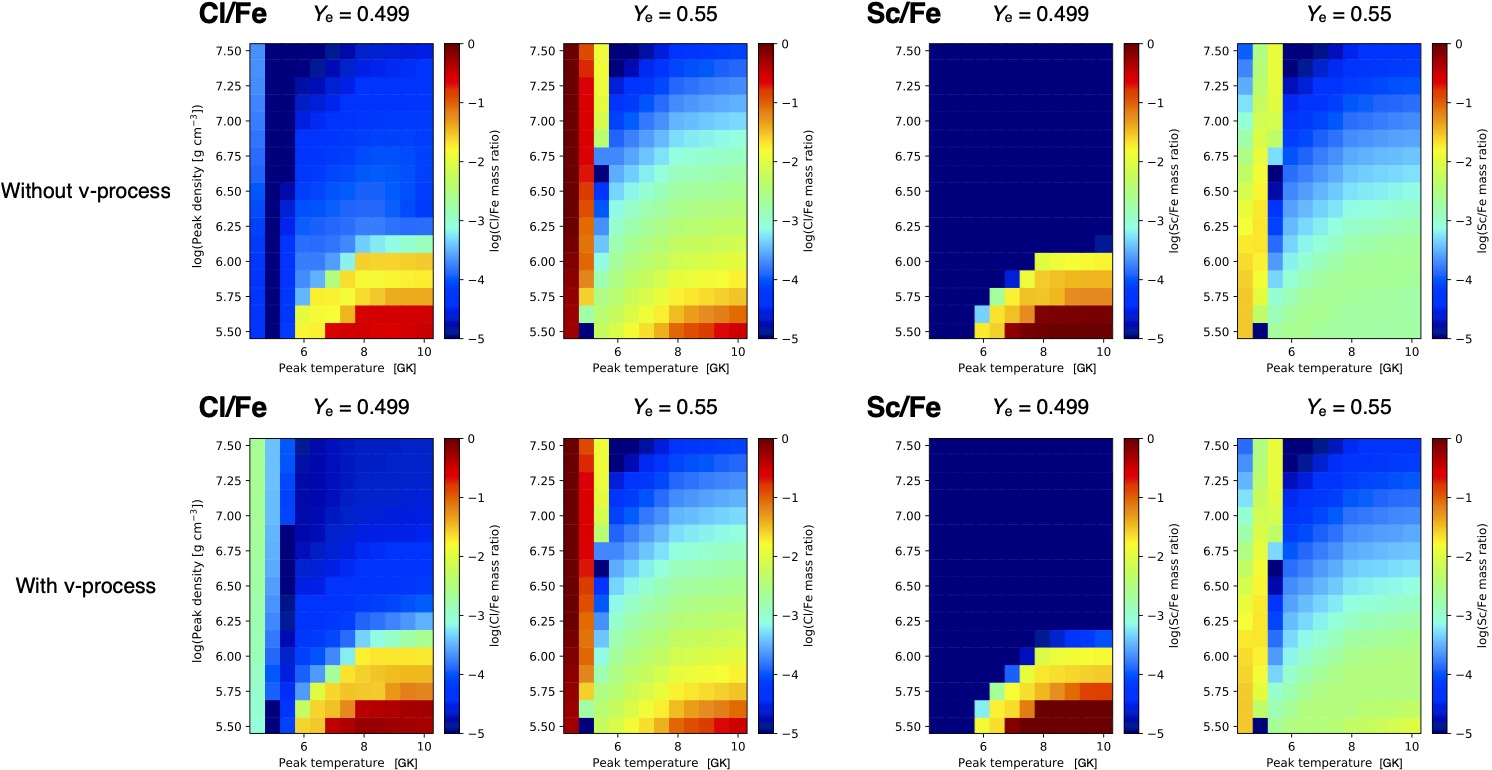}
 \end{center}
\caption{The Cl/Fe and Sc/Fe mass ratios in the peak temperature-density plane with (bottom row) and without (top row) the $\nu$-process. Electron fraction values of $Y_{\rm e} =$ 0.499 and 0.55 are assumed in these calculations. Here, we used the thermodynamic trajectories taken from our 1D supernova model. All the stable isotopes are included in the results. In the incomplete Si burning region ($T_{\rm peak} < 5$ GK), the Cl/Fe mass ratio is high and affected by the $\nu$-process, however the Sc/Fe mass ratio is low even when the $\nu$-process is included. As a common property, both the Cl/Fe and Sc/Fe mass ratios are significantly increased by strongly neutrino-processed proton-rich environment, with or without the $\nu$-process. In the calculations, we assume neutrino temperatures of $T_{\nu_{\rm e}} = T_{\bar{\nu}_{\rm e}} =$ 4 MeV/$k$ and $T_{\nu_{\rm x}} =$ 6 MeV/$k$ and a total neutrino energy of $3\times10^{53}$ erg.}
\label{fig:f5}
\end{figure*}

We have demonstrated that the observation of the Fe-rich ejecta in the Cassiopeia A supernova remnant can be a unique tool to probe the neutrino-matter interactions during the explosion. In particular, we have proposed that the observation of Mn produced in the innermost region of the supernova explosion could provide strong evidence of neutrino nucleosynthesis (both the $\nu$- and $\nu p$-processes). In our research, it has been found that the amount of Mn without the $\nu$- and $\nu p$-processes is too small to observe it, suggesting that our measured Mn/Fe mass ratio of 0.14--0.67\% supports the presence of significant neutrino interactions in the Cassiopeia A supernova. If the observed Mn/Fe mass ratio purely reflects the production at the innermost region of the supernova, this could be the first firm confirmation of the neutrino-matter interactions in an individual supernova.

The simulations in this study are based on 1D SN models, whereas actual SN are inherently 3D. Still our 1D framework is quite valuable for bridging the gap between the computationally more challenging 3D simulations and the observational data.  For example, we could adapt thermodynamic trajectories from 3D simulations into our work and explore the effects of neutrino interactions. This would then help guide future 3D models in the quest to produce realistic explosions.  Some multi-dimensional SN studies find evidence for anisotropic neutrino radiation \citep[e.g.,][]{2004ApJ...608..391K,2006A&A...457..281B,2008ApJ...685.1069O,2009ApJ...694..664M,2009ApJ...704..951K,2012A&A...537A..63M}, which could be tested by observation by mapping the distribution of elements sensitive to the neutrino processes (such as Mn) in supernova remnants.  This will require the high sensitivity of the next-generation X-ray observatories.

\begin{figure*}[t!]
 \begin{center}
  \includegraphics[bb=0 0 1586 634, width=16cm]{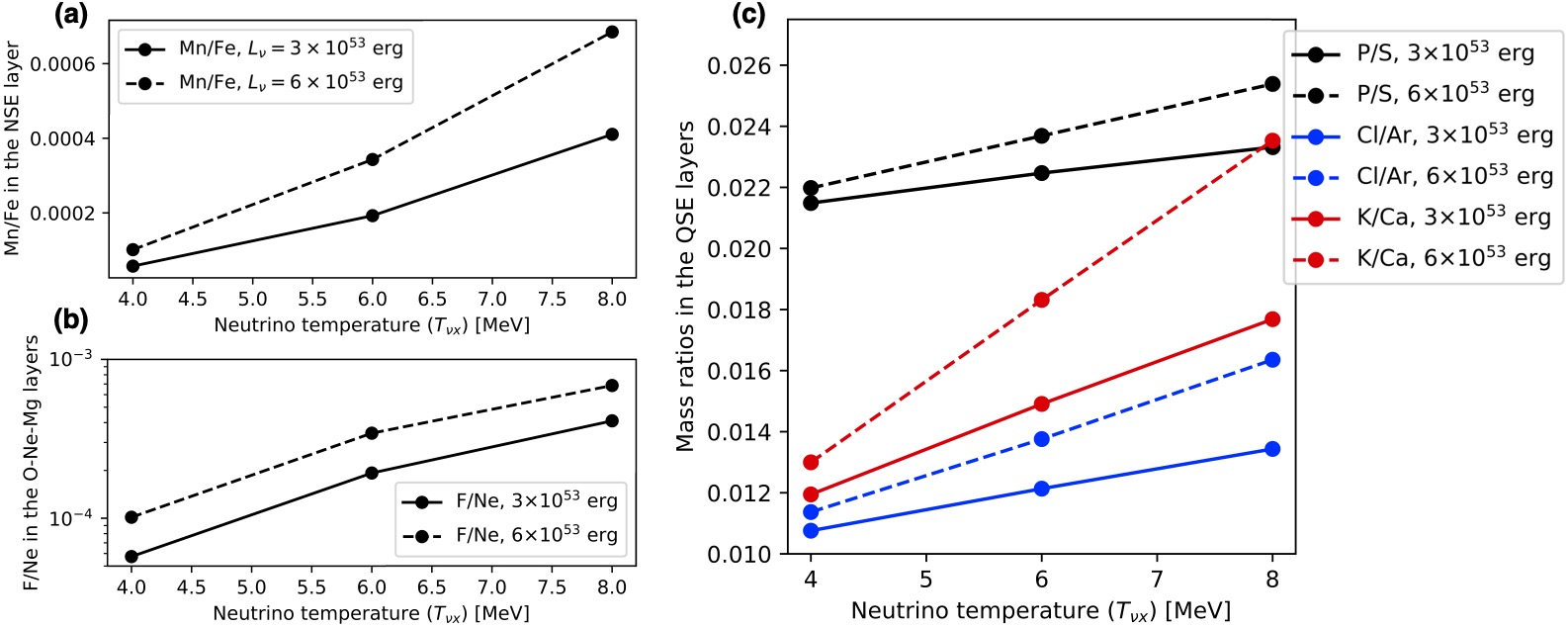}
 \end{center}
\caption{The dependence of elemental mass ratios on neutrino parameters in different nucleosynthetic layers of a core collapse SN. Here we assume three cases for the neutrino temperatures: (1) $T_{\nu_{\rm e}} =$ 3.4 MeV/$k$, $T_{\bar{\nu}_{\rm e}} = T_{\nu_{\rm x}} =$ 4 MeV/$k$ as a low temperature case, (2) $T_{\nu_{\rm e}} = T_{\bar{\nu}_{\rm e}} =$ 4 MeV/$k$, $T_{\nu_{\rm x}} =$ 6 MeV/$k$ as an intermediate temperature case, (3) $T_{\nu_{\rm e}} =$ 4 MeV/$k$, $T_{\bar{\nu}_{\rm e}} =$ 5 MeV/$k$, $T_{\nu_{\rm x}} =$ 8 MeV/$k$ as a high temperature case. (a) The Mn/Fe mass ratios in the NSE layer with the different neutrino parameters. The Mn abundance is enhanced by the reaction $^{56}{\rm Ni}(\nu,\nu^\prime p)^{55}{\rm Co}$. (b) The F/Ne mass ratios in the O/Ne layer with the different neutrino parameters. The F abundance is enhanced by the reaction $^{20}{\rm Ne}(\nu,\nu^\prime p)^{19}{\rm F}$. (c) The P/S, Cl/Ar, K/Ca mass ratios in the QSE layer with the different neutrino parameters. The P, Cl and K abundances are enhanced by the reactions $^{32}{\rm S}(\nu,\nu^\prime p)^{31}{\rm P}$, $^{36}{\rm Ar}(\nu,\nu^\prime p)^{35}{\rm Cl}$ and $^{40}{\rm Ca}(\nu,\nu^\prime p)^{39}{\rm K}$, respectively. The charged-current reactions also contribute to increase the  abundances.}
\label{fig:f-temp}
\end{figure*}

Future observations with X-ray calorimeters \citep[e.g.,][]{2018SPIE10699E..22T,2018SPIE10699E..1GB,2019JATIS...5b1001G} will provide further information on neutrino-processed ejecta. For example, in the $\nu$-process, the production of other elements is also enhanced by neutrino-induced reactions. The proton-rich ejecta are also expected to have enhanced odd-Z elements \citep{2018ApJ...852...40W}; however, current X-ray detectors do not have the
spectral resolution to search for these low abundance species. To observe these rare elements will provide us more fruitful discussions on the supernova neutrino physics. In particular, we hope that following questions can be answered in the next few decades.
\begin{itemize}
  \item Was the Fe-rich structure synthesized in the proton-rich or neutron-rich region? Or, is it a mixture of both?
  \item What are the total neutrino luminosity and neutrino temperature in core-collapse supernovae?
  \item Is the supernova neutrino radiation anisotropic?
\end{itemize}
We here discuss what kind of observations will lead to these solutions as a conclusion to this paper.

The first question will be the easiest to be answered. In the proton-rich ejecta, the abundances of specific elements (especially odd-Z elements) are high. In Figure \ref{fig:f5}, we show the Cl/Fe and Sc/Fe mass ratios, which could be the most promising means of answering the question, calculated in the peak temperature-density plane. These elements are significantly enhanced by the complete Si burning regime ($T_{\rm peak} > $ 5.5 GK) in the proton-rich environment, irrespective of additional $\nu$-processes being included or not. Therefore, if the amounts of these elements could be measured, we could test the nucleosynthesis in such a proton-rich environment. They are only measurable with a high resolution non-dispersive spectrometer like an X-ray calorimeter, since the current X-ray CCD data do not allow these emission lines to be separated from other emission lines or to be distinguished from the continuum emission.

As for the second question, some elements enhanced by the $\nu$-process, such as Mn, will be helpful as already discussed in section \ref{sec:mn2}. Figure \ref{fig:f-temp} shows the dependence of some elemental mass ratios on neutrino parameters, which is useful to discuss the potential to limit the neutrino parameters. Mn at the innermost Fe-rich region of core-collapse supernovae is one of the most sensitive elements for the $\nu$-process (Figure \ref{fig:f-temp} (a), the nuclear statistical equilibrium: NSE layer), where the abundance varies more than a factor of a few depending on the neutrino parameters. We could investigate it in other remnants, although there do not seem to be many other core-collapse supernova remnants that show Fe-rich structures like Cassiopeia A. For example, G350.1-0.3 and G292.0+1.8 exhibit asymmetrically distributed Fe ejecta \citep[e.g.,][]{2008ApJ...680L..37G,2011ApJ...731...70L,2020ApJ...905L..19B,2021ApJ...912..131T,2019ApJ...872...31B}, which will be other targets to test the neutrino emissions during the explosion. However, it would be difficult to discuss elemental compositions purely within the $\alpha$-rich freezeout region, because of the mixing with other regions in these remnants. As another tool to examine the $\nu$-process, fluorine could be a very useful element in the X-ray calorimeter era (Figure \ref{fig:f-temp} (b)). $^{19}$F is produced in the O- and Ne-rich region via $^{20}$Ne($\nu, \nu^\prime p$)$^{19}$F \citep[e.g.,][]{1990ApJ...356..272W,2003AJ....126.1305C,2011ApJ...739L..57K,2018ApJ...865..143S,2019ApJ...885..139G}. In the O/Ne-rich region, F is almost never synthesized without the neutrino process, making it a very sensitive element for the $\nu$-process. While it would be difficult in the case of Cassiopeia A and G350.1-0.3, which show strong absorption by the interstellar medium, it may be possible to limit neutrino emissions by measuring the F/Ne ratio for some core-collapse supernova remnants such as G292.0+1.8, 1E 0102.2-7219 and Puppis A, which have bright O and Ne emission lines \citep[e.g.,][]{2019ApJ...872...31B,2017A&A...597A..35P,2019ApJ...873...53A,2010ApJ...714.1725K}. Also, as seen in Figure \ref{fig:f-temp} (c), some other elements produced in the incomplete Si layer (i.e., QSE layer), such as P, Cl and K, also depend on the neutrino parameters, although there would be difficult to separate the neutrino nucleosynthesis from other effects (e.g., explosive and stellar nucleosynthesis).

While the last question is probably the most difficult to tackle, observing the spatial distribution of rare $\nu$-process elements could be key to addressing this issue. In particular, the distribution of F/Ne in O/Ne-rich supernova remnants, such as G292.0+1.8, 1E 0102.2-7219 and Puppis A, will be interesting to test anisotropic neutrino radiation. However, it will be difficult to reach a firm conclusion because of many uncertainties in both modeling and observation.

As discussed above, over the next decade or so, spatially-resolved, high-resolution spectroscopy will allow us to investigate those rare elements not only in Cassiopeia A, but also in other remnants. Thus, X-ray observations of supernova remnants will open a new window to probe neutrino physics during SN explosions.


\vspace{0.1cm}
\noindent
{\bf Acknowledgements}
This work was supported by the Japan Society for the Promotion of Science (JSPS) KAKENHI grant No. JP19K14739, JP18H05223, JP20H00174, JP20H00158, JP19H00693 and JP23K13128. TY is supported in part by the Grants-in-Aid for Scientific Research on Innovative Areas (JP20H05249) from the Ministry of Education, Culture, Sports, Science, and Technology, Japan. SN is partially supported by Program of RIKEN for Evolution of Matter in the Universe (r-EMU), and Theoretical and Mathematical Sciences Program of RIKEN (iTHEMS). JPH acknowledges support for X-ray studies of SNRs from NASA grant NNX15AK71G to Rutgers University.  JPH, the George A.\ and Margaret M.\ Downsbrough Professor of Astrophysics, acknowledges the Downsbrough heirs and the estate of George Atha Downsbrough for their support.

\appendix
\section{Detailed spatially-resolved analysis for the Fe-rich region of Cassiopeia A}
\label{sec:detail}

\begin{figure*}[h!]
 \begin{center}
  \includegraphics[width=16cm, bb=0 0 1472 482]{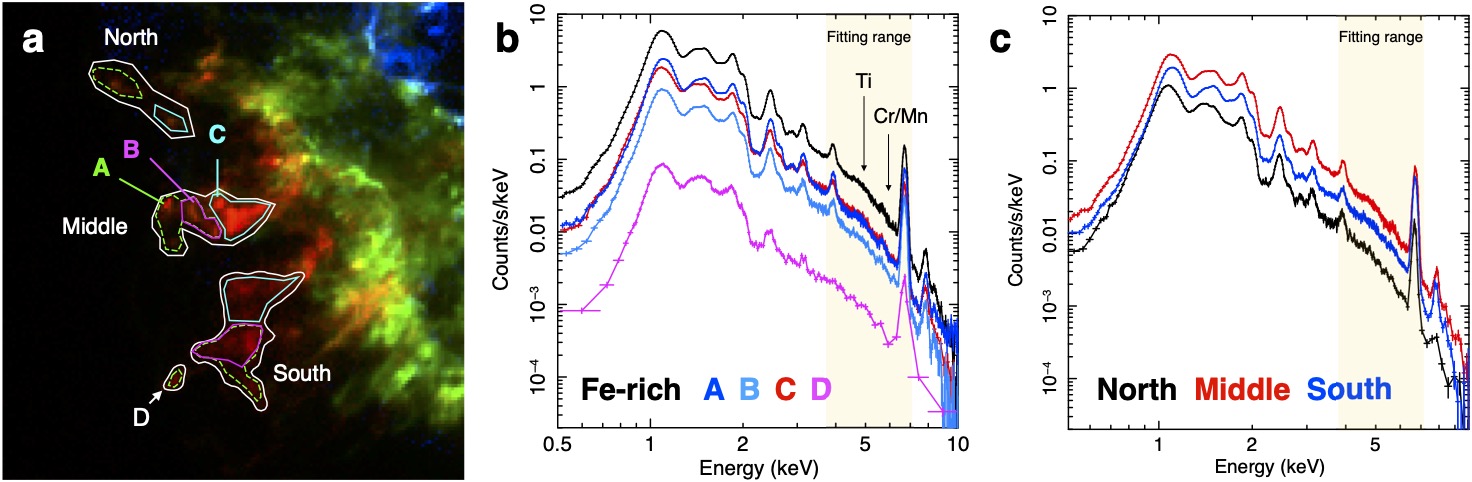}
 \end{center}
\caption{(a) X-ray image of Cassiopeia A around the southeastern Fe-rich region. Here we define new small regions (A, B, C, D, North, Middle, South) within the southeastern Fe-rich structure. (b,c) X-ray spectra extracted from the regions defined in (a). The spectral fittings were done using only the colored region (3.7--7.1 keV) in this figure.}
\label{fig:ap_f1}
\end{figure*}

\begin{table}[t!]
\caption{The best-fit parameters for the regions A, B, C, North, Middle and South in Fig.~\ref{fig:ap_f1}. The errors show 1$\sigma$ confidence level ($\Delta\chi^2$ = 1.0). The solar abundance in \cite{1989GeCoA..53..197A} is used.}
\begin{center}
\begin{tabular}{lrrrr}
\hline
Region ID.                              &  Fe-rich \citep{2020casATi}     &  A                             & B                                          & C \\ \hline
$n_{\rm H}$ (fixed) [cm$^{-2}$]         &   1.2$\times$10$^{22}$    &   1.3$\times$10$^{22}$                        & 1.3$\times$10$^{22}$                          & 1.3$\times$10$^{22}$\\
broadening $\sigma$ at 6 keV [keV]      &   (3.93$^{+0.08}_{-0.09}$)$\times$10$^{-2}$    &   (3.2$^{+0.3}_{-0.2}$)$\times$10$^{-2}$      & (3.2$^{+0.5}_{-0.7}$) $\times$10$^{-2}$       & (3.0$\pm$0.2)$\times$10$^{-2}$\\
Electron temperature $kT_{ e}$ [keV]    &   3.11$^{+0.10}_{-0.09}$     &   3.04$^{+0.17}_{-0.05}$                      & 3.1$^{+0.3}_{-0.2}$                           & 3.5$^{+0.2}_{-0.3}$\\
${\rm [Ca/H]/[Ca/H]}_{\odot}$           &   3.88$^{+0.05}_{-0.07}$     &   2.3$\pm$0.1                                 & 3.3$\pm$0.3                                   & 8.4$^{+0.2}_{-0.4}$\\
${\rm [Ti/H]/[Ti/H]}_{\odot}$           &   8.0$^{+1.3}_{-1.4}$ (5.7$\sigma$)    &   6.9$^{+2.8}_{-2.9}$ (2.4$\sigma$)                         & 4.6$^{+4.3}_{-3.2}$ (1.4$\sigma$)                           & 15.3$^{+4.6}_{-4.0}$ (3.8$\sigma$)\\
${\rm [Cr/H]/[Cr/H]}_{\odot}$           &   4.9$\pm$0.5 (9.8$\sigma$)     &   5.4$\pm$1.1 (4.9$\sigma$)                                 & 4.8$^{+1.4}_{-1.5}$ (3.2$\sigma$)                           & 9.6$^{+1.4}_{-1.0}$ (9.6$\sigma$)\\
${\rm [Mn/H]/[Mn/H]}_{\odot}$           &   6.6$\pm$1.3 (5.1$\sigma$)      &   7.1$\pm$2.7 (2.6$\sigma$)                                 & 12.5$^{+4.0}_{-4.8}$ (2.6$\sigma$)                          & 7.0$\pm$3.6 (1.9$\sigma$)\\
${\rm [Fe/H]/[Fe/H]}_{\odot}$           &   8.9$^{+0.2}_{-0.3}$    &   7.1$^{+0.6}_{-0.5}$                         & 8.7$\pm$1.0                                   & 16.7$^{+0.8}_{-1.1}$\\
$n_{e}t_{\rm low}$  [cm$^{-3}$ s]       &   1.5$^{+0.3}_{-0.2}$ $\times$10$^{10}$     &   0.9$^{+0.5}_{-0.4}$ $\times$10$^{10}$       & 3.1$^{+2.8}_{-1.3}$ $\times$10$^{10}$         & 1.6$^{+0.5}_{-0.3}$ $\times$10$^{10}$\\
$n_{e}t_{\rm high}$ [cm$^{-3}$ s]       &   2.9$^{+0.2}_{-0.1}$ $\times$10$^{11}$    &   2.4$^{+0.4}_{-0.2}$ $\times$10$^{11}$       & 2.7$^{+1.6}_{-0.4}$ $\times$10$^{11}$         & 2.6$\pm$0.2 $\times$10$^{11}$\\
redshift                                &   $-$9.53$^{+0.23}_{-0.03} \times 10^{-3}$    &   $-$8.9$^{+0.4}_{-1.1} \times 10^{-3}$    & $-$8.7$^{+1.3}_{-0.9} \times 10^{-3}$      & $-$1.21$^{+0.30}_{-0.05} \times 10^{-2}$\\
norm [$\frac{10^{-14}}{4 \pi D^2} \int n_{e} n_{\rm H} dV$ cm$^{-5}$] 
                                        & 8.2$^{+0.1}_{-0.2}$ $\times$10$^{-3}$    & 3.46$^{+0.05}_{-0.03}$ $\times$10$^{-3}$         & 1.55$^{+0.08}_{-0.14}$ $\times$10$^{-3}$      & 1.78$^{+0.07}_{-0.05}$ $\times$10$^{-3}$\\
$M_{\rm Ti}/M_{\rm Fe}$               & (1.6$\pm$0.3) $\times 10^{-3}$ & (1.7$\pm$0.7) $\times 10^{-3}$         & (1.0$^{+0.7}_{-0.9}$) $\times 10^{-3}$        & (1.6$^{+0.5}_{-0.4}$) $\times 10^{-3}$\\
$M_{\rm Cr}/M_{\rm Fe}$               & (5.2$\pm$0.5) $\times 10^{-3}$ & (7.1$\pm$1.4) $\times 10^{-3}$         & (5.2$^{+1.5}_{-1.6}$) $\times 10^{-3}$        & (5.4$^{+0.8}_{-0.6}$) $\times 10^{-3}$\\
$M_{\rm Mn}/M_{\rm Fe}$               & (3.8$\pm$0.8) $\times 10^{-3}$ & (5.2$\pm$2.0) $\times 10^{-3}$         & (7.4$^{+2.3}_{-2.8}$) $\times 10^{-3}$        & (2.2$^{+1.1}_{-1.1}$) $\times 10^{-3}$\\
$\chi^2$/d.o.f.                       & 23.63/35      & 24.39/35                                      & 46.15/49                                      & 30.31/49\\ \hline
Region ID.                              &&  North                             & Middle                                          & South \\ \hline
$n_{\rm H}$ (fixed) [cm$^{-2}$]         &&   1.0$\times$10$^{22}$                        & 1.2$\times$10$^{22}$                          & 1.3$\times$10$^{22}$\\
broadening $\sigma$ at 6 keV [keV]      &&   (2.2$^{+0.7}_{-0.8}$)$\times$10$^{-2}$      & (3.2$\pm$0.2) $\times$10$^{-2}$       & (2.9$^{+0.4}_{-0.2}$)$\times$10$^{-2}$\\
Electron temperature $kT_{ e}$ [keV]    &&   2.4$^{+0.2}_{-0.1}$               & 2.85$^{+0.07}_{-0.05}$                           & 3.3$\pm$0.2\\
${\rm [Ca/H]/[Ca/H]}_{\odot}$           &&   2.8$\pm$0.2                                 & 4.6$\pm$0.1                                   & 2.6$\pm$0.2\\
${\rm [Ti/H]/[Ti/H]}_{\odot}$           &&   8.7$^{+5.2}_{-4.9}$ (1.8$\sigma$)           & 8.5$\pm$2.3 (3.7$\sigma$)                           & 4.7$^{+3.0}_{-2.8}$ (1.7$\sigma$)\\
${\rm [Cr/H]/[Cr/H]}_{\odot}$           &&   4.1$^{+2.2}_{-2.1}$ (2.0$\sigma$)                   & 7.2$\pm$0.9 (8.0$\sigma$)                           & 3.6$\pm$1.0 (3.6$\sigma$)\\
${\rm [Mn/H]/[Mn/H]}_{\odot}$           &&   9.9$^{+6.7}_{-3.3}$ (3.0$\sigma$)                   & 9.4$^{+2.5}_{-2.4}$ (3.9$\sigma$)                          & 4.4$\pm$2.7 (1.6$\sigma$)\\
${\rm [Fe/H]/[Fe/H]}_{\odot}$           &&   6.8$^{+0.7}_{-1.1}$                         & 9.4$^{+0.2}_{-0.4}$                           & 10.0$^{+0.9}_{-0.6}$\\
$n_{e}t_{\rm low}$  [cm$^{-3}$ s]       &&   1.5$^{+1.9}_{-1.0}$ $\times$10$^{10}$       & 0.6$\pm$0.2 $\times$10$^{10}$         & 2.6$^{+1.3}_{-0.8}$ $\times$10$^{10}$\\
$n_{e}t_{\rm high}$ [cm$^{-3}$ s]       &&   2.7$^{+0.7}_{-0.3}$ $\times$10$^{11}$       & 2.6$\pm$0.1 $\times$10$^{11}$         & 2.6$^{+0.5}_{-0.3}$ $\times$10$^{11}$\\
redshift                                &&   $-$6.5$^{+1.6}_{-0.6} \times 10^{-3}$    & $(-$1.24$\pm0.02) \times 10^{-2}$      & $-$8.9$^{+0.6}_{-0.4} \times 10^{-3}$\\
norm [$\frac{10^{-14}}{4 \pi D^2} \int n_{e} n_{\rm H} dV$ cm$^{-5}$] 
                                        && 1.47$^{+0.09}_{-0.10}$ $\times$10$^{-3}$         & 5.29$^{+0.09}_{-0.06}$ $\times$10$^{-3}$      & 2.69$^{+0.10}_{-0.06}$ $\times$10$^{-3}$\\
$M_{\rm Ti}/M_{\rm Fe}$               && (2.3$^{+1.4}_{-1.3}$) $\times 10^{-3}$         & (1.6$\pm$0.4) $\times 10^{-3}$        & (0.8$\pm$0.5) $\times 10^{-3}$\\
$M_{\rm Cr}/M_{\rm Fe}$               && (5.6$^{+3.0}_{-2.9}$) $\times 10^{-3}$         & (7.1$\pm$0.9) $\times 10^{-3}$        & (3.4$\pm$0.9) $\times 10^{-3}$\\
$M_{\rm Mn}/M_{\rm Fe}$               && (7.5$^{+5.1}_{-2.5}$) $\times 10^{-3}$         & (5.2$\pm$1.3) $\times 10^{-3}$        & (2.3$\pm$1.4) $\times 10^{-3}$\\
$\chi^2$/d.o.f.                       && 37.40/35                                      & 44.29/35                                      & 39.26/49\\
\hline
\end{tabular}
\label{tab_FeRich}
\end{center}
\end{table}

\begin{figure*}[h!]
 \begin{center}
  \includegraphics[width=14cm, bb=0 0 1380 572]{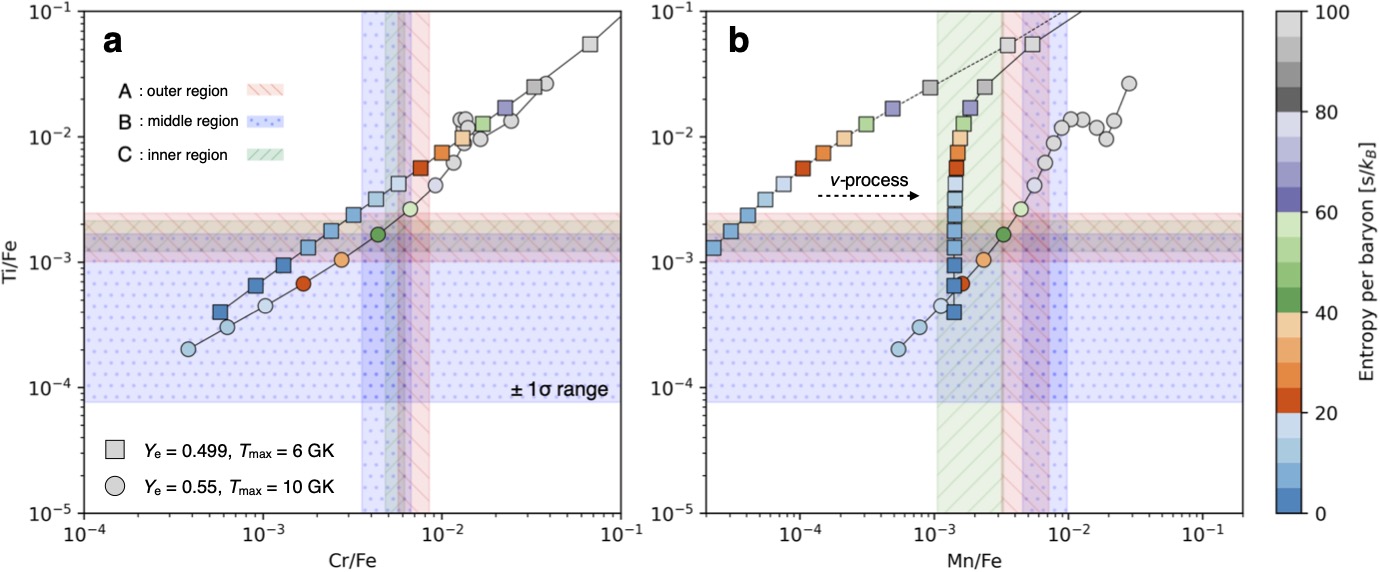}
  \includegraphics[width=14cm, bb=0 0 1384 576]{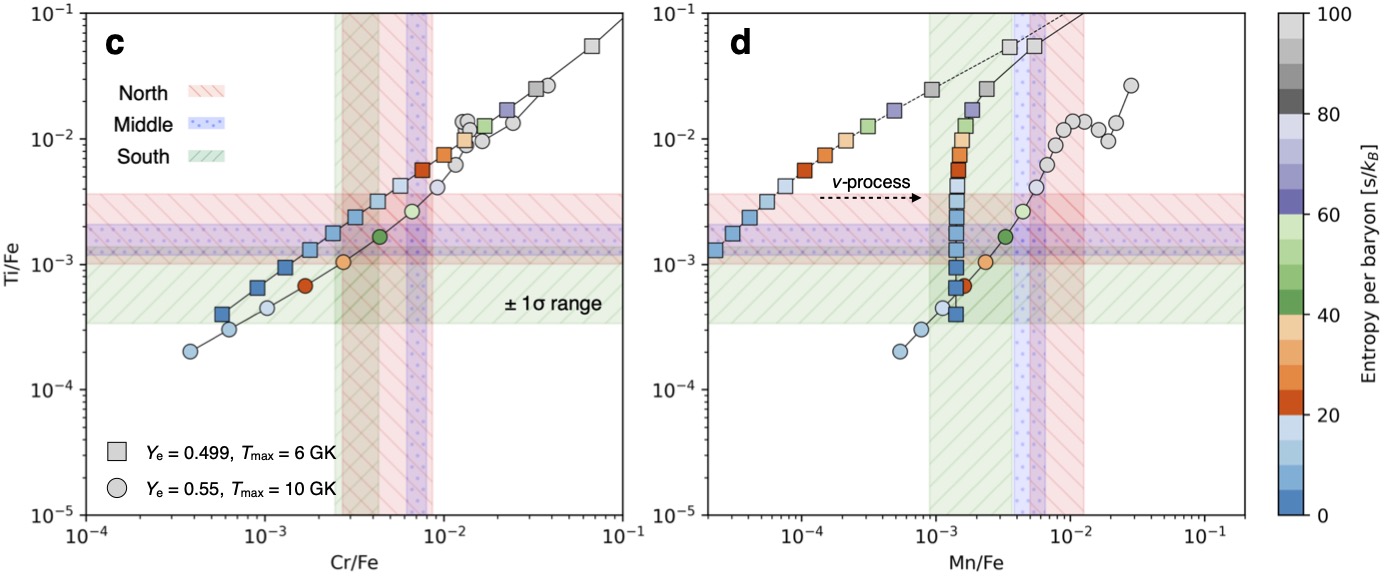}
 \end{center}
\caption{As same as Figure \ref{fig:f1}, but the colored areas show the mass ratios measured in the fine-detailed regions.}
\label{fig:ap_f4}
\end{figure*}

While the number of X-ray photons in the southeastern Fe-rich region are limited, the elemental abundance measurements in more fine-detailed regions would be useful to discuss the differences in physical parameters at each local region. Thus, in this section, we perform such spectral analyses. In Figure \ref{fig:ap_f1}a, we defined several small Fe-rich regions (A, B, C, D, North, Middle, South) among which we investigate differences in elemental abundances. The spectra (Figure \ref{fig:ap_f1}b,c) show that contamination from Si-rich regions is comparable in all regions (see the K$\alpha$ lines from Si, S, Ar and Ca), although the D knot at the outer edge of the Fe-rich structure has a little less contamination. If we can observe D knot with Athena or Lynx in the future, we will be able to more purely discuss the effects of neutrinos in terms of its elemental composition. 


In Table \ref{tab_FeRich}, we summarize the results of the spectral analysis for the fine-detailed regions (excluding the D knot). Following the same approach as in section \ref{sec:obs}, we used the X-ray spectra only in the 3.7--7.1 keV band for each region and fitted with an absorbed thermal plasma model with a gsmooth model that is used for modelling the lines broadened by thermal broadening and/or multiple velocity component. As a result, we found that all the regions have super-solar metallicities for the important elements (Ti, Cr, Mn) in our research, although the statistical errors are significantly larger than that in the entire Fe-rich region (i.e., the white contour region in Figure \ref{fig:ap_f1}a). The statistical significance of each line is above 1.4 $\sigma$ in all the fine-detailed regions. The Middle and C regions have the brightest Fe-rich structure (see Figure \ref{fig:ap_f1}a) and therefore have the highest significance levels for the line detection. While the statistical errors are large, we find here that the observed mass ratios may differ from region to region.

Based on the spectral analysis, we compared the observed Ti/Fe, Cr/Fe and Mn/Fe mass ratios with those in our theoretical calculations (Figure \ref{fig:ap_f4}). The mass ratios in different regions seem to have different acceptable models, which may reflect different physical processes. If it is real, future deep and precise X-ray observations will be useful to understand the cause of the diversity within these regions.

\section{OBSERVATIONS WITH XMM-NEWTON AND SUZAKU}
\label{sec:other}

\begin{figure*}[h!]
 \begin{center}
  \includegraphics[bb=0 0 1518 616, width=16cm]{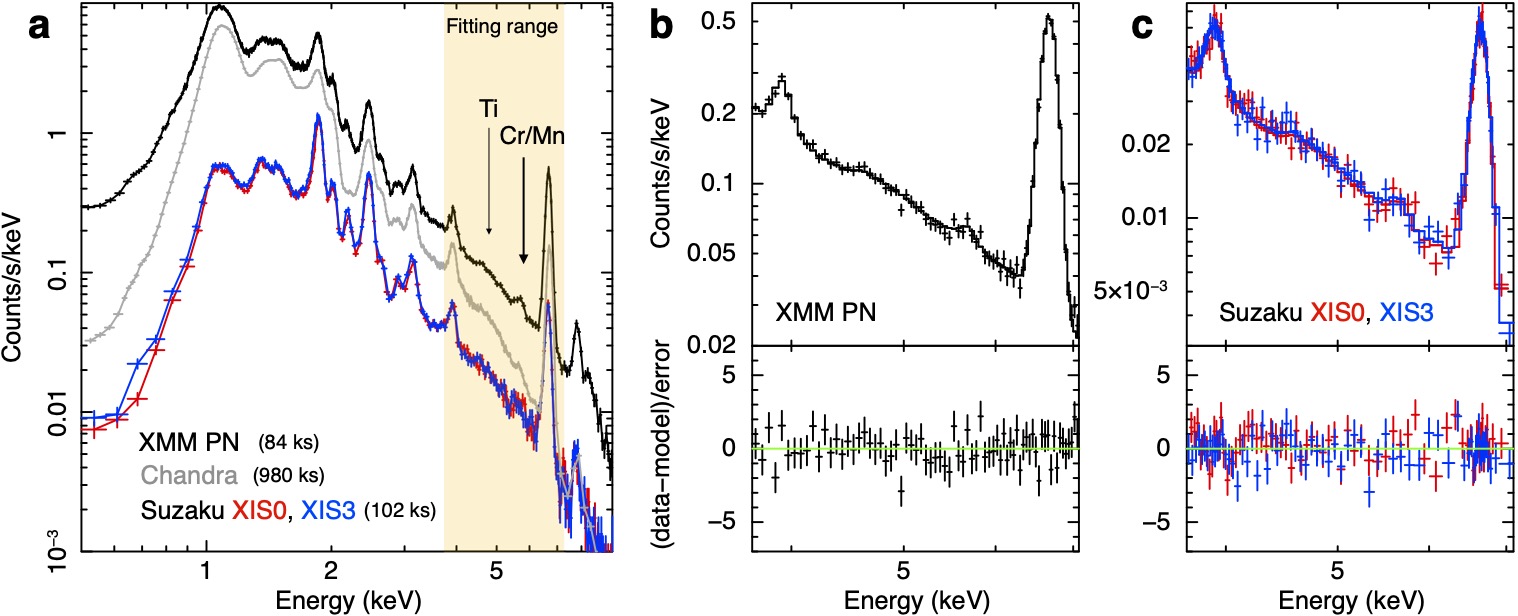}
 \end{center}
\caption{(a) Comparison of X-ray spectra in the Fe-rich region with XMM-Newton, Chandra and Suzaku. (b) X-ray spectrum in 3.7--7.1 keV with the XMM PN. (c) X-ray spectrum in 3.7--7.1 keV with the Suzaku XIS0 and XIS3.}
\label{fig:ap_f2}
\end{figure*}

To check the consistency among observations with different satellites, we analyzed X-ray observations of Cassiopeia A by XMM-Newton and Suzaku. For the analysis of the XMM-Newton observations, we used all the XMM PN data covering the Fe-rich region (Obs.ID 0097610501, 0097610801, 0110010301, 0110011001, 0110010201, 0412180101, 0110011801). Here, the XMM MOS data covering the Fe-rich region were limited due to each observation mode and pointing position, thus we did not use it. The data reduction was performed in a standard way using SAS version 20.0.0, where the total exposure time was 83.9 ks after the reduction. In the case of Suzaku, we analyzed the deepest observation in 2012 (Obs.ID 507038010). The total counts in 3.7--7.1 keV are $\sim$13\% (XMM PN) and $\sim$5\% (Suzaku XIS0+XIS3) of the Chandra data. In the analysis for both XMM-Newton and Suzaku, we used the Fe-rich region that shown as the white contour in Figure \ref{fig:ap_f1}a for extracting X-ray spectra.

\begin{table}[t]
\caption{The best-fit parameters with XMM-Newton and Suzaku. The errors show 1$\sigma$ confidence level ($\Delta\chi^2$ = 1.0). The solar abundance in \cite{1989GeCoA..53..197A} is used.}
\begin{center}
\begin{tabular}{lrrrr}
\hline
                                        &  XMM-Newton                                   & Suzaku                                 \\ \hline
$n_{\rm H}$ (fixed) [cm$^{-2}$]         &   1.2$\times$10$^{22}$                        & 1.2$\times$10$^{22}$               \\
broadening $\sigma$ at 6 keV [keV]      &   (2.2$^{+0.7}_{-0.4}$)$\times$10$^{-2}$      & ---   \\
Electron temperature $kT_{ e}$ [keV]    &   3.0$^{+0.2}_{-0.1}$                         & 2.5$^{+0.2}_{-0.1}$              \\
${\rm [Ca/H]/[Ca/H]}_{\odot}$           &   2.5$\pm$0.2                                 & 3.1$^{+0.3}_{-0.2}$                \\
${\rm [Ti/H]/[Ti/H]}_{\odot}$           &   5.9$^{+3.8}_{-3.4}$ (1.7$\sigma$)           & 5.0$^{+4.8}_{-4.7}$ (1.0$\sigma$)           \\
${\rm [Cr/H]/[Cr/H]}_{\odot}$           &   4.3$^{+1.3}_{-1.2}$ (3.6$\sigma$)           & 4.7$\pm$1.6  (2.9$\sigma$)          \\
${\rm [Mn/H]/[Mn/H]}_{\odot}$           &   $<$ 3.9 ($<$ 1$\sigma$)                     & 4.6$\pm$4.0 (1.1$\sigma$)            \\
${\rm [Fe/H]/[Fe/H]}_{\odot}$           &   6.1$^{+0.7}_{-0.3}$                         & 3.5$^{+0.3}_{-0.4}$                 \\
$n_{e}t_{\rm low}$  [cm$^{-3}$ s]       &   1.6$^{+0.7}_{-0.5}$ $\times$10$^{10}$       & 2.4$^{+25.1}_{-2.3}$ $\times$10$^{8}$ \\
$n_{e}t_{\rm high}$ [cm$^{-3}$ s]       &   (1.7$\pm$0.1) $\times$10$^{11}$             & 2.7$^{+0.4}_{-0.1}$ $\times$10$^{11}$ \\
redshift                                &   ($-$7.4$^{+0.3}_{-0.6}) \times 10^{-3}$     & ($-$2.6$^{+1.1}_{-0.8}) \times 10^{-3}$  \\
norm [$\frac{10^{-14}}{4 \pi D^2} \int n_{e} n_{\rm H} dV$ cm$^{-5}$] 
                                        & 1.0$^{+0.6}_{-0.3}$ $\times$10$^{-2}$         & 1.3$\pm$0.1 $\times$10$^{-2}$\\
$M_{\rm Ti}/M_{\rm Fe}$                 & (1.4$\pm$1.0) $\times 10^{-3}$                & (2.6$^{+2.5}_{-2.4}$) $\times 10^{-3}$ \\
$M_{\rm Cr}/M_{\rm Fe}$                 & (6.5$^{+1.9}_{-1.8}$) $\times 10^{-3}$        & (12.6$\pm$4.2) $\times 10^{-3}$ \\
$M_{\rm Mn}/M_{\rm Fe}$                 & $<$ 3.3 $\times 10^{-3}$                      & (6.8$^{+6.0}_{-5.9}$) $\times 10^{-3}$ \\
$\chi^2$/d.o.f.                         & 71.41/61                                      & 139.93/124                     \\
\hline
\end{tabular}
\label{tab_FeRich3}
\end{center}
\end{table}

Figure \ref{fig:ap_f2}a shows the comparison of X-ray spectra among Chandra, XMM-Newton and Suzaku. The strong Si and S emission lines suggest that the X-ray spectrum by Suzaku is highly contaminated from other regions due to its worse angular resolution \citep[whose half power width is $\sim$1.9;][]{2007PASJ...59S...1M}. On the other hand, the X-ray spectrum by XMM-Newton is quite similar to that by Chandra, suggesting that its angular resolution of $\sim$15$^{\prime\prime}$ \citep{2001A&A...365L...1J} is good enough to study the Fe-rich structure. Following the same approach as in section \ref{sec:obs}, we used the X-ray spectra only in the 3.7--7.1 keV band for  the spectral analysis and fitted with an absorbed thermal plasma model as shown in Figure \ref{fig:ap_f2}b,c. We find that the observed Ti/Fe, Cr/Fe and Mn/Fe by XMM-Newton and Suzaku agree with those by Chandra within the error ranges, although only upper limits (1-$\sigma$) were set in some cases due to poor photon statistics. A deep XMM-Newton observation that is $\sim$10 times longer than the current ones will be possible to make a discussion equivalent to the current Chandra data.

\section{Model Dependency}
\label{sec:model}

\begin{figure*}[h!]
 \begin{center}
  \includegraphics[bb=0 0 1088 862, width=16cm]{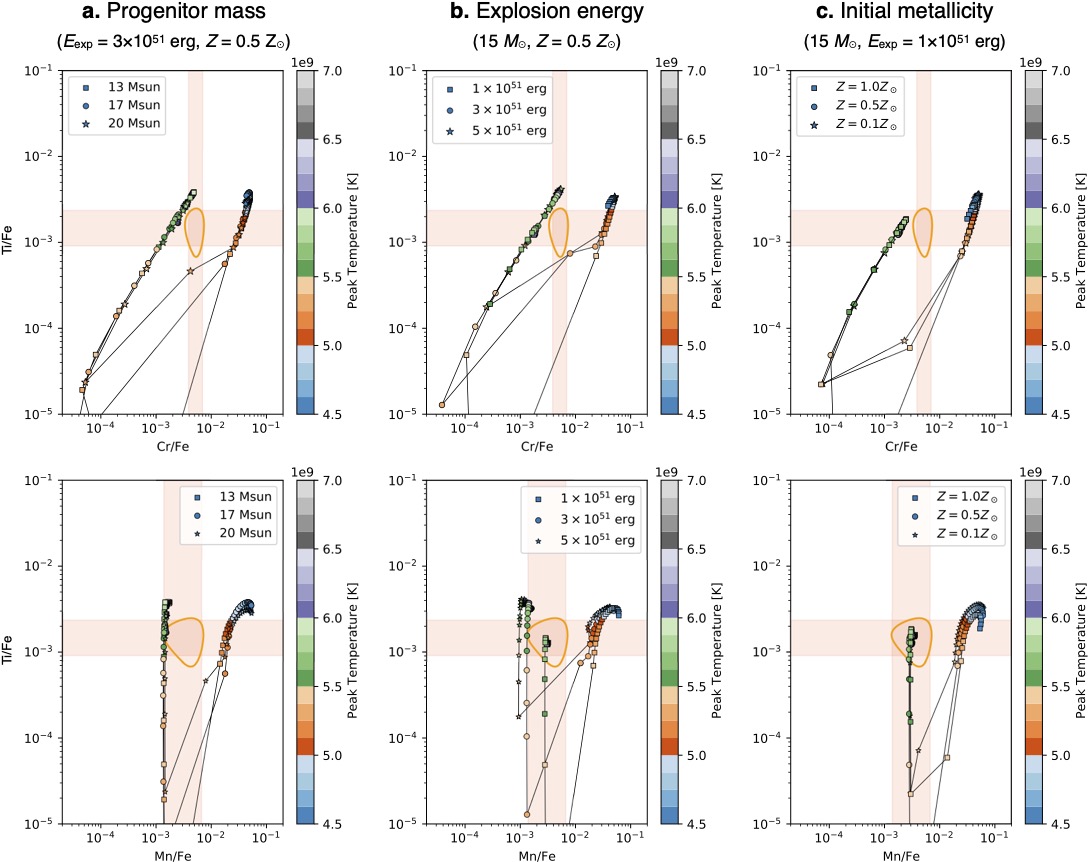}
 \end{center}
\caption{The observed Ti/Fe, Cr/Fe and Mn/Fe ratios compared with supernova models that have different parameters.}
\label{fig:ap_f5}
\end{figure*}

In sections \ref{sec:mn} and \ref{sec:mn2}, we compared the observed mass ratios with some theoretical calculations based on 1D supernova models. Here we discuss how the model parameters affect the elemental ratios used in our study.

In Figure \ref{fig:ap_f5}, we compare the observed mass ratios with several supernova models that have different parameters (i.e., progenitor mass, explosion energy and initial metallicity). Interestingly, the results show that the model trajectories for the Ti/Fe and Cr/Fe ratios, which are sensitive to the entropy during the explosion, are nearly identical for all models. The most important parameter to determine the amount of these elements is the explosion energy, where higher energies produce larger yields at the innermost region because a stronger realization of $\alpha$-rich freezeout. Also in the case of Mn/Fe, the explosion energy changes the ratio significantly. In our calculations, we determined the location of the mass cut so that the ejected $^{56}$Ni mass is 0.07 $M_\odot$ \citep[see][]{2020ApJ...893...49S}. Therefore, if we assume a smaller explosion energy, the mass-cut position moves inward and the amount of Mn by the $\nu$-process increases. For example, the model with the lower explosion energy of $1\times$10$^{51}$ erg shows a better agreement with the observation. On the other hand, the progenitor mass and initial metallicity do not change these mass ratios so much. These results support that our conclusion does not depend on the model parameters in the case of 1D calculations.

Ideally, it would be better to compare the observations with multi-dimensional models since supernova explosions are multi-dimensional phenomena. In particular, the Fe-rich structure focused on in this study must be related to multi-dimensional effects during the explosions. On the other hand, the current multi-dimensional calculations of supernovae still contain many uncertainties for explaining real supernovae. We hope future supernova modeling and observations will lead to a more accurate understanding.

\bibliography{sample631}{}
\bibliographystyle{aasjournal}

\end{document}